\documentstyle[psfig]{l-aa}

\def\ros{{\sl ROSAT}}
\def\etal{{et\,al.}}

\def\grad{$^\circ$}

\def\degs{\ifmmode ^{\circ}\else$^{\circ}$\fi}
\def\amin{\ifmmode ^{\prime}\else$^{\prime}$\fi}
\def\asec{\ifmmode ^{\prime\prime}\else$^{\prime\prime}$\fi}
\def\fss{\hbox{$.\!\!^{\rm s}$}}        % Fractions of seconds
\def\fdg{\hbox{$.\!\!^\circ$}}          % Fractions of degrees
\def\h{$^{\rm h}$}\def\m{$^{\rm m}$}

\newbox\grsign \setbox\grsign=\hbox{$>$}
\newdimen\grdimen \grdimen=\ht\grsign
\newbox\laxbox \newbox\gaxbox
\setbox\gaxbox=\hbox{\raise.5ex\hbox{$>$}\llap
     {\lower.5ex\hbox{$\sim$}}}\ht1=\grdimen\dp1=0pt
\setbox\laxbox=\hbox{\raise.5ex\hbox{$<$}\llap
     {\lower.5ex\hbox{$\sim$}}}\ht2=\grdimen\dp2=0pt

\font\pr=cmr5
\def\rxj{RX\,J0757.0+6306}

\begin{document}

   \thesaurus{06         % A&A Section 6: Form. struct. and evolut. of stars
              (02.01.2;  % accretion
%               08.02.2;  % binaries: eclipsing
               08.09.2;  % Stars: individual
               08.13.1;  % Stars: magn. fields
               08.14.2;  % Stars: CVs
               13.25.5)} % X-ray: stars

\title{A New Cataclysmic Variable RX\,J0757.0+6306: \\ 
     Candidate for the Shortest Period Intermediate Polar}

\author{
     G.H. Tovmassian\inst{1}
\and J. Greiner\inst{2}\thanks{Present address: Astrophysical Institute
        Potsdam, An der Sternwarte 16, 14482 Potsdam, Germany}
\and P. Kroll\inst{3}
\and P. Szkody\inst{4}
\and P.A. Mason\inst{5}
\and F.-J. Zickgraf\inst{2, 6}\thanks{Present address: Observatoire de
         Strasbourg, 11 rue de l'Universit\'e, F-67000 Strasbourg, France}
\and J. Krautter\inst{6}
\and I. Thiering\inst{6}
\and A. Serrano\inst{7}
\and S. Howell\inst{8}
\and D.R. Ciardi\inst{8}
}

\institute{Instituto de Astronom\'{\i}a, UNAM, Ensenada, B.C., M\'{e}xico
\and MPI f\"ur  extraterrestrische Physik, 85740 Garching, Germany
\and Sternwarte Sonneberg, 96515 Sonneberg, Germany
\and University of  Washington, Seattle, Washington, USA
\and NMSU, Las Cruces, New Mexico, USA
\and Landessternwarte K\"onigstuhl, Heidelberg, Germany
\and INAOE, Puebla, Mexico 
\and University of Wyoming, Laramie, Wyoming, USA}

   \date{Received 18 February 1998 / Accepted 19 March 1998}

   \maketitle

   \markboth{G.H. Tovmassian \etal}{RX J0757.0+6306}

   \begin{abstract}

A new cataclysmic variable is identified as the optical counterpart of the
faint and hard
X-ray source RX\,J0757.0+6306 discovered  during the ROSAT all-sky survey.
Strong double-peaked emission lines 
bear evidence of an accretion disc via an S--wave which varies
with a period of 81$\pm 5$~min. We identify this period as the orbital 
period of the binary system. CCD  photometry 
reveals an additional period of 8.52$\pm 0.15$ min. which was stable over 
four nights.
We  suggest that \rxj\ is  possibly an intermediate polar, but we cannot 
exclude the possibility that it is a member of the  SU UMa group of dwarf 
novae.
\end{abstract}

\section{Introduction}

Within a project for the optical identification of a complete sample of 674
northern ROSAT All-Sky Survey (RASS) X-ray sources (which is a collaboration
between the Max-Planck-Institut f\"ur extraterrestrische Physik, Garching, the
Landessternwarte Heidelberg, Germany, and the
Instituto Nacional de Astrof\'isica, Optica y Electronica, Mexico)
several new cataclysmic variables were identified.
A detailed description of the project is given by Zickgraf \etal\ (1997).
The full catalogue with all identifications is published in Appenzeller 
\etal\ (1998).
Here we report the identification of the RASS X-ray source RX\,J0757.0+6306 
(= 1RXS J075700.5+630602). 

Cataclysmic variables (CVs) are close binary systems with a white dwarf 
primary  accreting matter supplied by a late type main-sequence 
secondary star via an accretion disc or along magnetic field lines of the 
white dwarf. Magnetic CVs, where 
the white dwarf has a sufficiently strong magnetic field to affect the 
accretion trajectory, form two distinctive subclasses: the high-field polars, 
and the low-field intermediate polars (IPs). These subclasses are 
characterized by well-defined observational properties (Cropper 1990;
Patterson 1994; Warner 1995).
The polars are usually soft X-ray emitters and have near synchronously 
rotating 
WD, the IPs are harder X-ray sources and show a second periodicity due to the 
asynchronously rotating WD. In some cases a third period is observable, 
which is interpreted as the beat period between orbital  and 
spin periods. 

Besides differences in the flux distribution and variability, the orbital 
period distribution of the various subclasses of CVs were also noticed to 
be different (Kolb 1995). Polars tend to cluster below the period gap 
(2 h $< P_{orb}<$ 3 h), while IPs 
are preferentially above the gap. Non-magnetic CVs are distributed almost 
equally. All subclasses however show deficiency of systems in the period gap 
and a short-period cutoff at $\rm P_{min}=80$ min (the minimum period). 
The statistically significant properties of the period distribution are 
presumed  to have  an evolutionary origin (Verbunt \& Zwaan 1981, 
Verbunt 1984, King 1988, Kolb and Ritter 1992). The rapidly increasing 
number of 
magnetic CVs discovered from the  ROSAT data has a significant impact on the 
above mentioned distribution and its consequences. 

\section{Observations}

\rxj\ was scanned in the RASS between Sep. 28--30, 1990
for a total exposure time of 420 sec. \rxj\ is found as a source with 
a total of 55 photons, which corresponds to a vignetting corrected
count rate of 0.13$\pm$0.02 cts/s. 
No strong variability in the X-ray intensity is seen at this level.
The spectrum, as derived from these 55 photons is rather hard, extending
up to 2.4 keV (the upper bound of the PSPC) as evidenced by the hardness 
ratios HR1 = 0.80$\pm$0.08 and HR2 = 0.42$\pm$0.13.
The best fit X-ray position is (equinox 2000.0): RA = 07\h 57\m 00\fss3,
Decl. =  +63\degr 05\amin 56\asec\   ({b\sc ii}=+31\fdg3).

\rxj\ was observed by the Extreme Ultraviolet Explorer Satellite
(EUVE) on 3 Oct 1997 UT. A 50 ksec observation was performed and data
collected in both the EUV spectrographs ($70-800 \AA$ wavelength coverage)
and in the Deep Survey Instrument (a EUV imager with peak sensitivity
near $90 \AA$). The source was not detected in either detector.

The original optical identification observations were carried out at the 
2.1\,m telescope operated by INAOE at Cananea, Sonora, Mexico. 
The optical counterpart of the X-ray source was identified with an 
emission line star using a low-resolution multiobject spectrograph. 
Further detailed study of the object, and its identification as a new CV,
was done at the 2.1\,m telescope of the Observatorio Astron\'omico Nacional 
de San Pedro M\'artir, Mexico. The Boller \& Chivens 
Spectrograph with a 600 l/mm grating was 
used to obtain spectra in the $3600 - 5700 \AA$ range with 
$ 4.5 \AA$ FWHM resolution.  
Exposure times of 300 sec were  chosen in order to be able 
to derive the orbital period of the system. Later we observed the object  
alternating between $4200-6300 \AA$ and $4600-6700 \AA$ wavelength ranges 
with the same spectral and time resolution.

\begin{figure}[t]
      \vbox{\psfig{figure=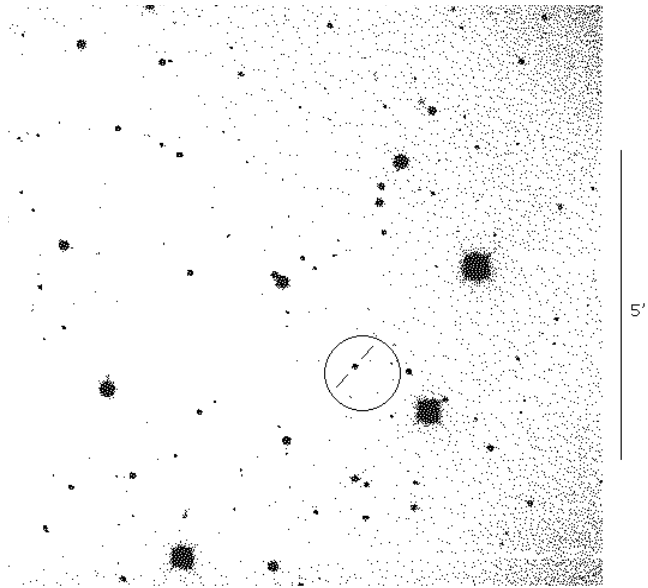,width=8.8cm,%
          bbllx=1.4cm,bblly=5.1cm,bburx=20.7cm,bbury=23.cm,clip=}}\par
      \vspace{-0.5cm}
      \caption[fchart]{Image of the field around the X-ray source
          RX\,J0757.0+6306 (centroid position with a 40\arcsec\, (3$\sigma$)
          error circle). The cataclysmic variable is marked by two dashes.}
      \label{fc}
\end{figure}

We obtained eight spectra at the Astrophysical Research Consortium (ARC) 
3.5-m telescope at Apache Point Observatory.
The double-beam spectrograph (DIS) was employed in high-resolution (3$\AA$) 
mode set in two wavelength regions $4200-5100 \AA$ and $7900-8900 \AA$. Due to 
weather problems we covered only about one orbital period with a 
time-resolution of 8 min.

Optical photometry was performed during four nights in April 1996
at the Sonneberg Observatory 600/1800 mm reflector, equipped with a
385$\times$578 pixel EEV CCD. Exposure times were 40--60 sec with overall time 
resolution of about 80--120 sec using a Johnson B filter. In February 1997, 
one more night of photometric observations were performed with the same 
settings.

Additional optical photometry was performed at the Red Buttes
Observatory of the University of Wyoming. The object was observed 
on the nights of 9 and 10 Feb 1997 UT using the RBO 24" telescope equipped 
with ST-6 CCD. All integration times were 150 seconds and a "clear" filter was
employed in order to obtain the highest possible time resolution.
The "clear" filter is simply using the ST-6 in an unfiltered mode, thus
the bandpass approximates the QE curve of the CCD itself, similar to a
broad R+I bandpass. 

Infrared observations were made using the University of Wyoming 2.3\,m
IR telescope (WIRO) on the night of 09 Feb 1997. The observations were
made using the Aerospace camera, a LN2 cooled Nicmos IR array imaging camera.
The JHK observations
were 30 sec integrations over an 90 min period.  The filters were rotated
so successive filter observations are about 2.5 min apart.

The combined log of all optical observations is presented in Table~\ref{log}.

\begin{table*}[th]
\vspace{-0.25cm}
\caption{Log of Optical Observations}
\vspace{-0.2cm}
\begin{tabular}{ccccrrl}
      \noalign{\smallskip}
      \hline
      \noalign{\smallskip}
Date & JD & Telescope + Equip. & Filter        & Duration & Exp. & Site \\
     &    &                    & Wavelength    &  min.     & sec. &     \\ 
 \noalign{\smallskip}
 \hline
 \noalign{\smallskip}
 1996 April 10    & 2450183 & 2.1m, B\&Ch sp. & 3600--5700 & 120 & 300 & SPM\\
 1997 February 09 & 2450489 & 3.5m, DIS sp. & 4200--5200 & 70 & 480 & APO\\
 1997 February 09 & 2450489 & 3.5m, DIS sp. & 7900--8900 & 70 & 480 & APO\\
 1997 March 02    & 2450509 & 2.1m, B\&Ch sp. & 4200--6300 & 180 & 300 & SPM\\
 1997 March 03    & 2450510 & 2.1m, B\&Ch sp. & 4600--6700 & 90 & 300 & SPM\\
 1996 April 16    & 2450190 & 0.6m, CCD       & B & 150 & 40 & Sonneberg\\
 1996 April 17    & 2450191 & 0.6m, CCD       & B & 150 & 40 & Sonneberg\\
 1996 April 18    & 2450192 & 0.6m, CCD       & B & 70 & 40 & Sonneberg\\
 1996 April 20    & 2450194 & 0.6m, CCD       & B & 155 & 60 & Sonneberg\\
 1997 February 01 & 2450481 & 0.6m, CCD       & B & 225   & 60   & Sonneberg\\
 1997 February 09 & 2450489 & 0.6m, CCD       & R+I & 120 & 150 & RBO\\
 1997 February 09 & 2450489 & 2.3m, NICMOS    & J,H,K & 90 & 30 & WIRO\\
 1997 February 10 & 2450490 & 0.6m, CCD       & R+I & 120 & 150 & RBO\\
\noalign{\smallskip}
      \hline
    \end{tabular}
   \label{log}
   \end{table*}

The spectra were reduced using standard IRAF routines. 
The optimal extraction
 method was used to retrieve the spectra. Wavelength calibration was done using
 He-Ar arc spectra at the beginning and end of each set of spectra except for 
ARC spectra which were calibrated using a single arc spectrum obtained prior 
to the object. Flux calibration was possible, although slit width limitations 
and non-alignment to the parallactic angle makes these results less reliable.

\section{Results}

\subsection{Identification and position}

The sequence of spectra of the emission line object taken with the 2.1 m 
telescope shows strong emission lines of the Balmer series, He{\sc i} and He{\sc ii}
on top of a blue continuum. It is the brightest and only object reachable for 
spectrophotometry in the 40\arcsec\ error box of the ROSAT ASS.
We measured the position of the optical counterpart of \rxj\ as (equinox 2000.0)
R.A. = 07\h57\m00\fss5, Dec. = 63\degr06\arcmin02\arcsec ($\pm$1\arcsec).
Fig. \ref{fc} shows a finding chart with the  cataclysmic variable 
marked.

\subsection {Spectroscopy}

\begin{figure}[t]
  \vbox{\psfig{figure=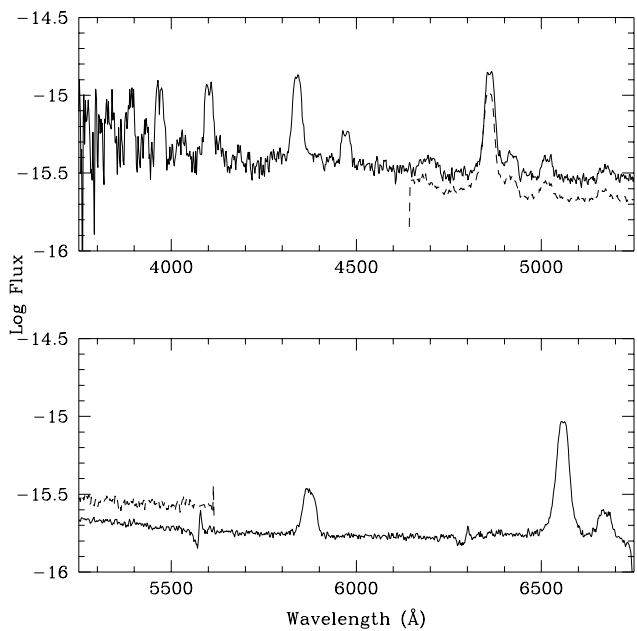,width=8.8cm}}
   \caption[sp]{The optical spectrum of RX\,J0757.0+6306. Two integrated 
      spectra from different epochs are presented.  }
   \label{sp}
   \end{figure}

A typical Balmer series accompanied by He{\sc i} and He{\sc ii} lines in emission (see Fig~\ref{sp})
identifies the magnetic CV nature of \rxj\ fairly well .  
Details of the spectra are given in  Tab~\ref{log}.  The spectra from each night are summed and flux calibrated. The two channels overlap around H$\beta$ and have continuum fluxes which differ by no more than 0.2 mag. 

The emission lines are double-peaked due to the contribution from a strong 
accretion disc. In addition, the spike corresponding to the S--wave component 
is clearly visible at most  phases, moving back and forth inside the lines.
We used the double-Gaussian deconvolution method suggested by Schneider and 
Young (1980), and Shafter (1985) in order to measure radial
velocity variations. The method is especially designed to measure line wing
variations by varying the separation and FWHM of double Gaussians, thus 
taking into account different parts of the wings. A wide range of Gaussian 
half--separations (250 -- 2000 km/sec) was used.  However, reasonable 
radial velocity curves were obtained only in the narrow range between 450 to 
650 km/sec, which actually corresponds to the central parts of the line, 
where the S--wave dominates. This method assumes that the orbital period is 
known well enough to measure precise radial velocities in the application of 
diagnostics, as described in the above mentioned papers.  
Instead, since the period was not known, we used the radial velocity 
measurements for the period search.

\begin{figure}
   \vbox{\psfig{figure=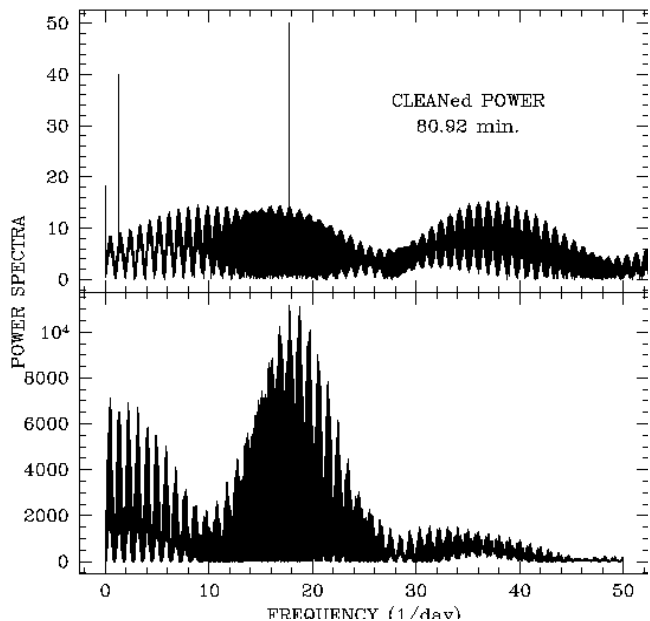,width=8.6cm,%
      bbllx=0.7cm,bblly=5.7cm,bburx=20.0cm,bbury=24.4cm,clip=}}
   \vspace{-0.5cm}
    \caption[sp]{The power spectrum of radial velocity measurements of  
                 RX\,J0757.0+6306.  }
   \label{powsp}
   \end{figure}

It is known from the study of various systems
that the S--wave is caused  by the bright spot on the outher edge of the disc, 
where inflowing matter strikes the disc (Smak 1976; Young \etal\ 1981;
Shafter \& Szkody 1984). WZ Sge is the classical example of such a system 
(Hack \& la Dous, 1993; Kaitchuk \etal\ 1994). The orbital motion is often 
seen as a radial velocity variation of the outer parts of the line wings. 
However, the amplitude of these variations is smaller than that of the bright 
spot. Usually the radial velocity curve of the primary is phase shifted 
relative to the S--wave by almost 180\grad. We barely can see the line 
variations corresponding to the WD primary. We fitted single Gaussians with 
centers corresponding to the S--wave component with different widths and 
subtracted them from  actual line profiles, but still we were unable to see 
any substantial (measurable) period variations in the rest of the lines. 
However, the clear waveform RV curve emerged at all measured Balmer lines 
at a separation of double Gaussians between 450--650 km/sec. So we selected 
550 km/sec half-separation Gaussians for analysis because that choice produced
the smoothest waveform and it lies in the center of the range of waveform 
variations.

Our analysis involves mainly the measurements of H$\beta$. The H$\gamma$ 
and H$\alpha$ lines were also  measured and checked for consistency against 
H$\beta$.  The period search revealed equally strong (1 day aliased) peaks at 
86,  81 and 76 min  periods. A search on a
nightly basis tends to a higher frequencies than on the
combined data. The 76 min peak was dominant in the April 96 observations 
covering almost 1.5 orbital periods (Tovmassian \etal\ 1997).

\begin{figure}
    \vbox{\psfig{figure=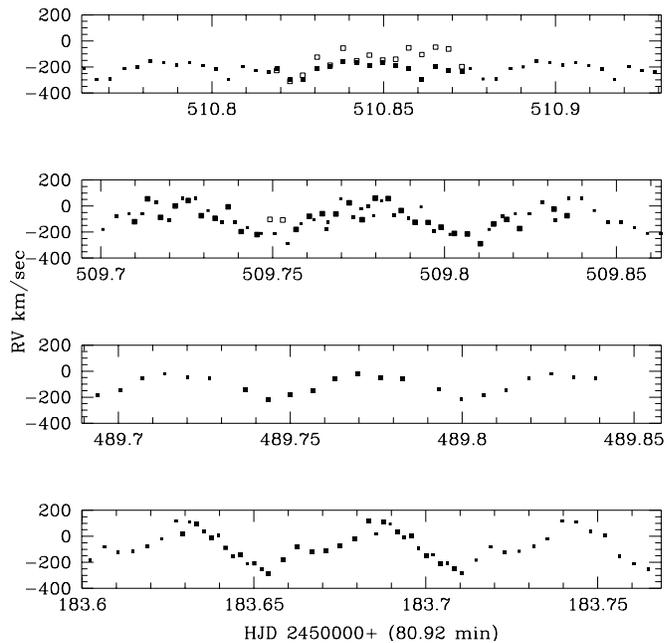,width=8.8cm,%
      bbllx=0.7cm,bblly=5.7cm,bburx=20.1cm,bbury=24.5cm,clip=}}
    \vspace{-.3cm}
    \caption[sp]{The radial velocity measurements folded by the best period. 
Each night is shown on separate panel. In the first three panels from the 
bottom the radial velocities of H$\beta$ are presented. In the third panel 
two points out of three rejected are shown by open symbols (see text), the 
third point drops out of the boundaries of the graph. In the fourth (top) 
panel measurements of H$\alpha$ are presented by the filled symbols, while 
open symbols are of H$\beta$. In all panels the points are repeated one phase 
shifted back and forth.  }
   \label{rvcurve}
   \end{figure}

Three measurements were rejected (from a total of 77) after thorough 
inspection of each spectrum
 in the combined set of data, because the line profiles were affected by
 cosmic rays. The combined power spectrum was also affected by the data
 corresponding to the fourth night when the quality of spectra were the
 poorest and the spectral range was centered toward H$\alpha$. Since we had
 more signal at H$\alpha$, we replaced the RV measurements of H$\beta$ on
 the fourth night by corresponding measurements of H$\alpha$. The resulting
 power spectrum shows competitive peaks at 81 and 76 min respectively, with
 81 min being slightly stronger. Next we performed the CLEAN algorithm 
(Roberts,
 Leh\'ar \& Dreher 1987) on the H$\beta$ measurements and on the combined
 H$\beta$ + H$\alpha$ data. Both resulted in a single strong peak around 
 81 min estimation. The CLEANed spectra are presented in the top panel of
 Fig~\ref{powsp}. In the bottom panel an unCLEANed power spectrum of
 data incorporating the H$\beta$ and H$\alpha$ is presented. The RV curves
 folded at 81 min period are shown (each night on the separate panel) in the
 Fig~\ref{rvcurve}. Larger symbols in  Figure~\ref{rvcurve} correspond to the 
actual measurements. Smaller symbols are actual measurements which have been 
shifted back and forth in phase by one for clarity. In the top panel, 
measurements of H$\alpha$ are designated by
 filled symbols while those of H$\beta$ are open.

\begin{figure}
   \vbox{\psfig{figure=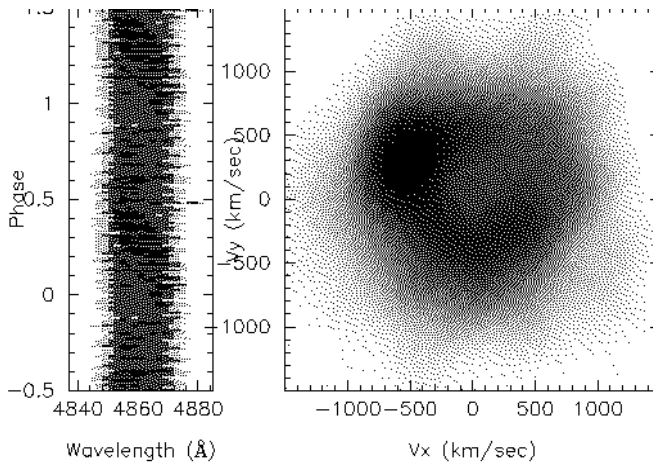,width=8.8cm,%
      bbllx=1.1cm,bblly=7.9cm,bburx=19.30cm,bbury=21.0cm,clip=}}
  \vspace{-.5cm}
    \caption[sp]{The velocity map of \rxj. The phase is chosen arbitrarily,
  so is the location of the hot spot on the figure seen as a darker spot on a 
  ring corresponding to the accretion disc.}
   \label{tomo}
   \end{figure}

The RV curve of Fig~\ref{rvcurve} looks smoother than the curves folded at 
aliased 
periods. We favor the 81 min period over the two others. Nevertheless, we 
cannot completely exclude the other periods as alternative solutions. 
Therefore, we must include a 5\,min uncertainty, although our
 estimation of the peak is precise to 0.014\,min. By comparison to similar 
systems, we assume that this period is the orbital period of the binary. 
However, the ephemerides of the stellar components remain unknown. We can also
assume that the system is at a very low inclination so that  we see it face on.
We applied the Doppler tomography method (backprojection method by Marsh and 
Horne 1988) to constrain the velocity map of the system. The H$\beta$ line 
from the first two nights were used. The 81 minute period was assumed to be 
the orbital period and phase was picked arbitrarily. Thus, the location of the 
dense spot on the velocity map (see Fig~\ref{tomo}) corresponding to the hot 
spot is also arbitrary, but it is evidence for the presence of a large 
accretion disc (donut structure)  with the prominent hot spot on its edge. 

We measured fluxes and equivalent widths of the majority of the visible lines. 
We checked the measured values (they are presented in Table \ref{lines}) 
with other CVs\footnote{according to Williams \& Ferguson (1982)}  since the 
system lies near the lower limit of orbital 
periods for hydrogen-composition cataclysmic variables. No abnormalities 
were detected. The  He {\sc i}  strength is within the limits for normal dwarf 
novae. The  He {\sc ii} 4686 \AA\ line is present, but its  strength does not 
indicate 
anything definite though in many CVs it often correlates with the magnetic 
field strength of the WD primary.

\begin{figure}
  \vbox{\psfig{figure=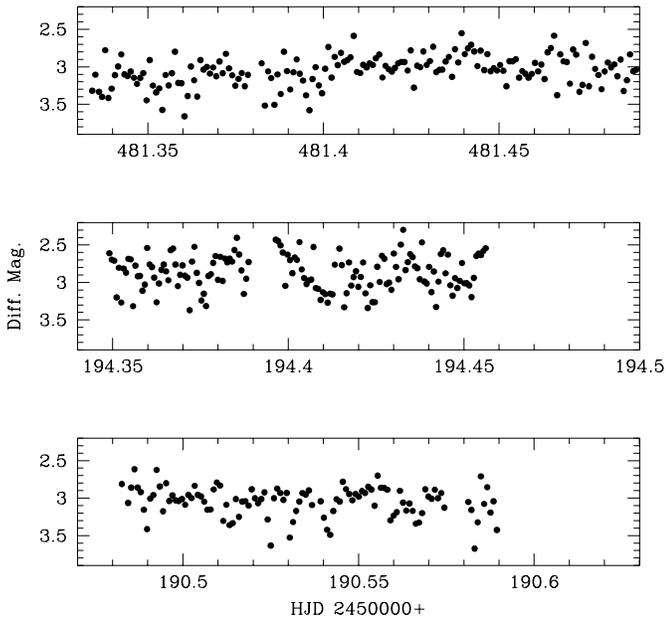,width=8.8cm,%
      bbllx=0.7cm,bblly=5.5cm,bburx=20.70cm,bbury=24.40cm,clip=}}
  \vspace{-0.5cm}
  \caption[lc]{Light curves of three separate nights. The x-axes have the 
       same scale.}
  \label{lc}
\end{figure}
 
\begin{table}
\caption{Measurements of spectral lines}
\begin{tabular}{lccll}
      \hline
      \noalign{\smallskip}
Emission & Equivalent& Flux($\times$10$^{-14}$)$\!\!$ & Relative  & 
      Range of  \\
 Line  &  Width  $\AA$  &  (erg/cm$^2$/s)  & Flux   &  Fluxes$\rm ^1$      \\ 
 \noalign{\smallskip}
 \hline
 \noalign{\smallskip}
 H$\alpha$    & -150(5) & $3.7$ & 103 & 91-188 \\
 H$\beta$     & -118(4) & $3.6$ & 100 & 100 \\
 H$\gamma$    & -90(4) & $3.4$ & 94 & 72--119 \\
 H$\delta$    & -70(3) & $2.7$ & 75 & 53--114 \\
 H$\epsilon$  & -45(4) & $1.9$ & 52 & -- \\
 He {\sc i} {\pr 5876}    & -38(4) & $0.65$     & 25 & 12--31\\
 He {\sc i} {\pr 5015}    & -14(2) & $0.37$     & 10 & 11--14\\
 He {\sc i} {\pr 4922}    & -18(3) & $0.55$     & 15 & 6--17\\
 He {\sc i} {\pr 4471}    & -21(1) & $0.72$     & 20 & 17--31 \\
 He {\sc i} {\pr 4026}    & -14(1) & $0.50$     & 15 & 9--20 \\
 He {\sc ii} {\pr 4686}   & -16.7(1.5) & $0.50$ & 14 & 15--34 \\
       \noalign{\smallskip}
      \hline
    \end{tabular}
   \vspace{-0.5cm}
   \label{lines}
   \end{table}

\subsection {Optical Photometry}

\begin{figure}
  \vbox{\psfig{figure=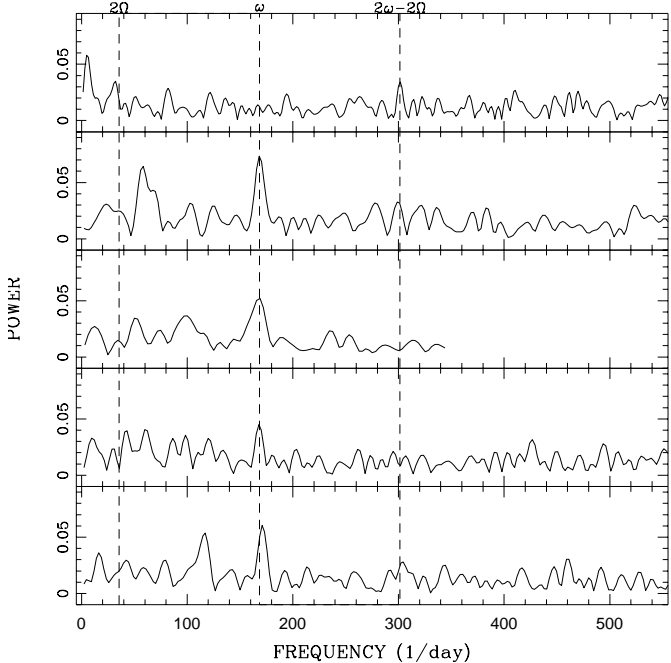,width=8.8cm,%
      bbllx=0.9cm,bblly=5.6cm,bburx=19.90cm,bbury=24.60cm,clip=}}
    \caption[sp]{The power spectra of each night of B photometry of  
    RX\,J0757.0+6306. The frequencies corresponding to the orbital 
    period, possible spin period and their harmonics are marked by 
   vertical lines.  }
   \label{phpowsp}
   \end{figure}

The light curves of the three longer CCD photometry runs are presented in 
Fig.~\ref{lc}.  The light curve demonstrates large amplitude flickering. 
However, no eclipses or modulations with the spectroscopic period 
were observed. This fits well with the spectroscopic results, namely that 
the system has a low inclination and a large accretion disk. 
Nevertheless, we conducted a period search on the photometric data. 
The period search on time series from separate nights revealed one 
repetitive peak around 169 cy/day at the first four nights. The 
amplitude varies from night to night, 
but the significance of the peak is almost the same on each night 
and does not exceed the noise level by more than 10--15\%. We 
combined the data of those nights and obtained similar results. The separate 
night peaks coincide with an accuracy of  $\pm$4~sec, while the precision of the determination is about 9\,sec. The fifth night of data, which is 
separated by $\approx 400$~day lag from the first four, does not show any 
signal at the 
above mentioned frequency. However there is a small peak which perfectly 
coincides with the $2(\omega - \Omega)$ frequency where $\Omega$ is the 
spectroscopic (orbital) period and $\omega$ is the frequency obtained 
on the previous nights. The power spectra of separate nights are presented 
in Figure \ref{phpowsp}. The first night is in the lower panel, the 
fifth night is in the top. The mentioned periods are indicated by vertical 
lines. We replaced the actual data from the  first four nights with random 
numbers spread exactly by the same range of magnitudes as the observed. We 
repeated the period search on the simulated random data and found no peaks 
at the 169 cy/day frequency or any other repetitive or significant signal. 
So we are confident that the peaks in the power spectra of the original 
data are  not caused by the temporal distribution of the exposures. 
The 169\,cy/day frequency corresponds to 8.52\,min, which constitutes about 
10\% of the orbital period. Although no conclusive dependence was found 
between the spin and orbital periods of intermediate polars, most of them 
tend to have spin periods of approximately 1/10 of the orbital. 

Thus, \rxj\ probably belongs to the family of intermediate polars. 
If confirmed, the  
ultra--short orbital period makes this system unique within this class. 

\subsection {Infrared Data}

The near infrared spectra which we obtained with the red chip of the DIS 
spectrograph at the ARC 3.5\,m telescope does not show any interesting 
features, particularly any absorption lines from the secondary star. That 
is not surprising, because the latter is expected to be a low--luminous 
late M dwarf while radiation of the system is strongly dominated by the disc. 
Extrapolating the dependence of the spectral type on the orbital period, 
presented by Patterson (1984), we estimate that the secondary
should be of spectral type M8 V for the orbital period of \rxj. We note
that there are a few exceptions in which the
period-secondary relation does not seem to work well for short period 
systems, and therefore types M4-6 could be possible too.

Through the use of IR standard stars, we found mean magnitudes
for \rxj\ of J=19.0$\pm$0.3 mag, H=18.5$\pm$0.3 mag, and 
K=18.1$\pm$0.3 mag.  An M8 V star should have an
absolute K magnitude of M$_K \sim 10$ mag according to Kirkpatrick \etal\ 
(1993).  Thus, if we assume that the
radius of the secondary is similar to that of a normal field dwarf and that
all of the K flux is contributed solely by the secondary, we find a distance
to \rxj\ of $D \sim 400$ pc. These assumptions are extreme (including that of 
the spectral type), but since they push the limits in opposite directions we 
find the estimate to be quite reasonable.

The weather was photometric for all these observations except that of 10
February 1997. Differential time-series light curves were produced for each
night/filter combination consisting of 2 hour runs in the optical and 
1.4 hour runs in the IR. 
The photometric behavior of \rxj\ in the IR showed constant
brightness light curves (within the errors) and no indication of any
short period (few minutes) oscillations. The near IR data showed
more modulation with an almost constant flickering-type behavior.
The February 10
data was interrupted for about half its length by clouds and is therefore
not very useful for short-period analysis. The  February 9 data showed
a constant value with flickering for about two-thirds of the time and
then a rapid increase (5 min) in brightness of 0.2 mags, but with no
change in the flickering amplitude. Period searches of the single short
February 9 photometric dataset 
yielded no conclusive periods.

\subsection{Archival photographic data}

We conducted a limited search in the all-sky patrol plates of
 Sonneberg Observatory. We looked through 194 photographic plates
 (blue sensitive) from the years 1958--1965 (these are in general the best
 plates available, because later air pollution degrades the sensitivity). 
We found one outburst of \rxj\ from these plates. It was seen on only one 
plate taken on 9 December 1964. Another plate one hour later was not as deep 
and the nearest other plates were taken on 9 November 1964  and 
9 February 1965 in which the object was  not visible,
 definitely being fainter than it was on 9 December 1964.

\section{Discussion}

A new cataclysmic variable is discovered with  interesting features:

\begin{enumerate}
\item The orbital period of $81\pm 5$ min puts \rxj\ near the hydrogen burning 
period minimum where CVs experience a turning point of their evolution.
Large flickering in the optical light curve and the observed
 spectral features of the object certainly show the presence of an accretion
 disc.

\item The limited search in the Sonneberg all-sky patrol plates revealed that 
 the system undergoes outburst activity. Another outburst was recorded
 (vsnet-alert No. 1379) shortly after the object's discovery was announced
 through the VSNET (vsnet-chat No 662). From the plate statistics we 
 can assume that the system has rather
 frequent outbursts. The amplitude of the outbursts of about 4 mag are
 typical for dwarf novae systems, but not as large as in SU~UMa superoutbursts
or the so called 
TOADs (tremendous outburst amplitude dwarf novae; Howell \etal\ 1995).

\item There are periodic light variations with a period of  8.5 min in the
light curve of the \rxj. We observed them directly on four  out
 of five occasions. In the fifth night a periodic signal with a side-band 
frequency was
 detected in the power spectrum.  Very recently, the 8.5\,min period was 
confirmed by R. Fried (vsnet-alert No 1387) from more prolonged observations.

\end{enumerate}

Thus, \rxj\ shows mixed characteristics, making its type classification
uncertain. From purely spectroscopic characteristics one may conclude that the 
new CV is a dwarf nova. Its short orbital period suggests that instead it may 
belong to the SU~UMa class or TOADs. But the repetitive
 detection of high-frequency pulses with a clearly fixed period indicates 
that it deserves a classification as an intermediate polar. This still needs 
to be confirmed by checking the coherency of the photometric pulses and by the 
detection of X-ray pulses. Intermediate polars  are CVs with the
 primary white dwarf rotating asynchronously due to its moderate magnetic
 field. Column accretion onto the magnetic poles results in the emission of 
high-energy radiation. This radiation is reprocessed elsewhere in the system 
into optical light which is modulated at periods shorter than the orbital 
period. The optical modulation can track the spin and/or the spin/orbit beat 
period of the binary (see the review by Patterson 1994).

The presence of X-ray emission in the quiescent state of  \rxj\ along with the 
moderate He~{\sc ii} 4686 \AA\
emission also argue in favor of a magnetic nature. The survey of 
non-magnetic CVs by van Teeseling \etal\ (1996) shows that the majority of 
X-ray emitting dwarf novae are of the SU UMa type, but they all are softer 
sources (HR1$\le 0.7$) than \rxj. There are a few long-period objects 
classified as non-SU UMa variables that are as hard as \rxj.  These belong 
to the VY~Scl, Z~Cam, and UX~UMa subclasses. We do not have any evidence 
which support a classification of 
\rxj\ as any of these types. Hence, since the rest of the CVs which are 
X-ray sources are magnetic, we conclude that \rxj\ is most probably magnetic.

Only two IPs have been detected with EUVE and only a few dwarf novae,
all of the latter during outburst. AM Herculis stars, particularly those
with high magnetic fields are  detected using EUVE. Thus, the
non-detection of \rxj\ with EUVE does not prove that it is not an IP.
It may indicate that \rxj\ could have a weak magnetic field (B $< 8$ MG), 
but given that only two IPs have EUVE detections and because of the rather
large distance of \rxj\ the non-detection is not 
considered unusual. On the other hand, the intensity of 
the He{\sc~ii} emission is not high enough to unambiguously classify it 
as a magnetic system. Silber (1992) set the following criteria for magnetic 
CVs: $20<$EW~(H$\beta)<40\AA$ 
and He{\sc ii}/H$\beta>0.4$. In our case, if the  larger equivalent width 
could be attributed to a shorter orbital period, the He{\sc ii}/H$\beta$ 
ratio is definitely below this criterium ($\approx~0.15$).

The existence of outbursts and the short orbital period of the system is  
in some discordance with the IP classification. Most IPs cluster above the 
2--3 hour period gap, while short period magnetic CVs are usually polars. 
However, a weak field IP will remain an IP even when it evolves towards 
shorter periods. In IPs, accretion outside of the Alfven radius remains in the 
form of a disc, while accretion inside the Alfven radius is dominated by flow 
along magnetic field lines. Outburst activity is uncommon since the inner 
part of the disc is disrupted by the magnetic field.

Nevertheless, neither outburst  activity nor short orbital period exclude the 
possibility of \rxj\ to be classified as an IP. In addition, the presence of 
a large disc in a short period magnetic CV suggests that the magnetic field is 
weak. Otherwise it would be a polar. For such a weak field  case it may not be 
surprising that \rxj\ appears to be an IP with some properties (i.e. outbursts)
similar to non-magnetic CVs, yet the evidence that it is a magnetic CV is 
compelling. Therefore, we offer \rxj\ as a candidate for the
shortest period intermediate polar.

\begin{acknowledgements}
GHT acknowledges support by grants from DGAPA IN109195 and CONACYT No 25454-E,
JG is supported by the Deutsche Agentur f\"ur Raumfahrtangelegenheiten (DLR)
GmbH under contract QQ 9602\,3 and PS acknowledges grants NSF AST9217911 
and NASA LTSA  NAG 53345.
The \ros\, project is supported by the German Bundes\-mini\-ste\-rium f\"ur
Bildung, Wissenschaft, Forschung und Technologie (BMBF/DLR) and the
Max-Planck-Society.
\end{acknowledgements}

\end{document}